\documentclass[12pt,a4paper]{article}
\usepackage[dvips]{graphicx,color}
\newcommand{\be}{\begin{equation}}
\newcommand{\ee}{\end{equation}}
\newcommand{\ba}{\begin{eqnarray}}
\newcommand{\ea}{\end{eqnarray}}
\newcommand{\baa}{\begin{eqnarray*}}
\newcommand{\eaa}{\end{eqnarray*}}

\newcommand{\dis}{\displaystyle}
\newcommand{\biq}{\mbox{\boldmath $q$}}
\newcommand{\bip}{\mbox{\boldmath $p$}}
\newcommand{\bif}{\mbox{\boldmath $f$}}
\newcommand{\biL}{\mbox{\boldmath $\Lambda$}}
\newcommand{\bil}{\mbox{\boldmath $\lambda$}}

\begin{document}
%\markboth{\LaTeXe{} Class for Lecture Notes in Computational
%Science and Engineering}{\LaTeXe{} Class for Lecture Notes in
%Computational Science and Engineering}
%\thispagestyle{empty}
%\title{Free-Energy Calculations in Protein Folding by 
%Generalized-Ensemble
%Algorithms}
%\author{Yuji Sugita\inst{1}$^,$\inst{2} \and Yuko Okamoto\inst{1}$^,$\inst{2}}
%\institute{Department of Theoretical Studies\\
%Institute for Molecular Science\\
%Okazaki, Aichi 444-8585, Japan
%\and
%Department of Functional Molecular Science\\
%The Graduate University for Advanced Studies\\
%Okazaki, Aichi 444-8585, Japan}
%\maketitle
%
{\pagestyle{empty}
\vskip 1.5cm

{\renewcommand{\thefootnote}{\fnsymbol{footnote}}
\centerline{\large \bf Free-Energy Calculations in Protein Folding}
\centerline{\large \bf by 
Generalized-Ensemble Algorithms}
}
\vskip 1.5cm
 
\centerline{Yuji Sugita\footnote{\ \ e-mail: sugita@ims.ac.jp}
and Yuko Okamoto\footnote{\ \ e-mail: okamotoy@ims.ac.jp}}
\vskip 1.0cm
\centerline{{\it Department of Theoretical Studies}}
\centerline{{\it Institute for Molecular Science}}
\centerline{{\it Okazaki, Aichi 444-8585, Japan}}
\centerline{and}
\centerline{{\it Department of Functional Molecular Science}}
\centerline{{\it The Graduate University for Advanced Studies}}
\centerline{{\it Okazaki, Aichi 444-8585, Japan}}

\vskip 1.0cm

%\centerline{Submitted to {\it Biopolymers (Peptide Science)}; 
%{\bf cond-mat/0012021}}
\centerline{Submitted to {\it Lecture Notes in Computational 
Science and Engineering}}

%\vskip 0.5cm

%\centerline{{\it {\bf Keywords:} protein folding; 
%generalized-ensemble algorithm; multicanonical algorithm;}}
%\centerline{{\it simulated tempering; replica-exchange method; parallel tempering}}

\medbreak
\vskip 1.0cm
 
\centerline{\bf ABSTRACT}
\vskip 0.3cm

%\begin{abstract}
We review uses of the generalized-ensemble 
algorithms for free-energy calculations in protein
folding.
Two of the well-known methods are 
multicanonical algorithm and 
replica-exchange method; the latter is also referred
to as parallel tempering.
We present a new generalized-ensemble algorithm
that combines the merits of the two methods; 
it is referred to as the replica-exchange 
multicanonical algorithm.
We also give a multidimensional extension of the
replica-exchange method.  Its realization as
an umbrella sampling method, which we refer to
as the replica-exchange umbrella sampling, is a 
powerful algorithm that can give free energy
in wide reaction coordinate space.

%\end{abstract}
%
%\vfill
%\newpage}
}

\section{Introduction}

Over the past three decades, a number of powerful
simulation algorithms have been introduced to the
protein folding problem (for reviews see, e.g.,
Refs.~\cite{RevSch}--\cite{RevHO2}).  For many
years, the emphasis has been placed on how to
find the global-minimum-energy conformation 
among a huge number of 
local-minimum states.  For complete understanding of
protein folding mechanism, however, the global
knowledge of the configurational space is
required, including the 
intermediate and denatured states of proteins.
For this purpose, free-energy calculations are
essential.

We have been advocating the uses of 
{\it generalized-ensemble algorithms} as the
methods that meet the above requirements
(for reviews see, e.g.,
Refs.~\cite{RevHO1,RevMSO}).
In this method
each state is weighted by a non-Boltzmann probability
weight factor so that
a random walk in potential energy space may be realized.
The random walk allows the simulation to escape from any
energy barrier and to sample much wider configurational
space than
by conventional methods.
Monitoring the energy in a single simulation run, one can
obtain not only
the global-minimum-energy state but also canonical ensemble
averages as functions of temperature by the single-histogram \cite{FS1}
and/or multiple-histogram \cite{FS2,WHAM} reweighting techniques
(an extension of the multiple-histogram method is 
also referred to as
{\it weighted histogram analysis method} (WHAM)
\cite{WHAM}).
  
Three of the most well-known generalized-ensemble methods are
perhaps {\it multicanonical algorithm} (MUCA) \cite{MUCA,MUCArev},
{\it simulated tempering} (ST) \cite{ST1,ST2}, and
{\it replica-exchange method} (REM)
\cite{RE1,RE2}.  (MUCA is also referred to 
as {\it entropic sampling} \cite{Lee,HSch} and 
{\it adaptive umbrella sampling} \cite{BK}.
ST is also referred to as the
{\it method of expanded ensemble} \cite{ST1}.
REM is also referred to as 
{\it parallel tempering} \cite{STrev}.  Details
of literature
about REM and related algorithms can be found in a
recent review \cite{IBArev}.)
Since MUCA was first introduced to protein
folding problem \cite{HO}, various generalized-ensemble
algorithms have been 
used in many applications in
protein and related systems (see Ref.~\cite{RevMSO}
and references therein).  In particular, free-energy
calculations in protein folding by generalized-ensemble
algorithms were explored in Refs.~\cite{HMO97,HOO}.
    
REM has been drawing much 
attention recently because the probability
weight factors are essentially known {\it a priori}, whereas
they are not in most of other generalized-ensemble algorithms
(and have to be determined by a tedius procedure).
In REM a number of
non-interacting copies (or replicas) of the 
original system
at different temperatures are
simulated independently and
simultaneously by the conventional Monte Carlo (MC) or 
molecular dynamics (MD) method. Every few steps,
pairs of replicas are exchanged with a specified 
transition
probability.

We have 
worked out the details for the replica-exchange
molecular dynamics algorithm \cite{SO}
(see also Ref.~\cite{H97}).
We have also developed a {\it multidimensional
replica-exchange method} (MREM) \cite{SKO} 
(see also Refs. \cite{Huk2,YP}).
In MREM we showed that REM 
is not limited
to tempering (or temperature
exchange) and that we can also 
exchange parameters in the potential
energy.
Namely, pairs of replicas with different temperatures
and/or different parameters of the potential energy
are exchanged during the simulation.
Important applications of MREM are
free-energy calculations.

The umbrella sampling method \cite{US} and free energy perturbation
method, which is a special case of umbrella sampling,
have been widely used to calculate the
free energies in chemical processes 
\cite{US} - \cite{SK2}. 
Although the effectiveness of the umbrella sampling method is
well known, its successful implementation requires a careful
fine tuning.  
Various generalizations of the umbrella sampling method have thus
been introduced to sample the potential energy surface more 
effectively.  The $\lambda$-$dynamics$ \cite{Ldyn1} - \cite{Ike} is such
an example, where the coupling parameter $\lambda$ is treated
as a dynamical variable.  Another example is the 
{\it multicanonical
WHAM} \cite{ONHN}, which combines the umbrella sampling with
multicanonical algorithm.
We have developed yet another generalization
of the umbrella sampling method (we refer to this method
as {\it replica-exchange umbrella sampling} (REUS)), 
which is based on the
multidimensional extension of the 
replica-exchange method \cite{SKO}.

REM is very effective and has already 
been used in many applications in protein 
systems (see Ref.~\cite{RevMSO} and references therein). 
However, REM also has a computational difficulty:
As the number of degrees of freedom of the system increases,
the required number of replicas also greatly increases, whereas 
only a single replica is simulated in MUCA or ST.
This demands a lot of computer power for complex systems.
Our solution to this problem is: Use REM for the weight
factor determinations of MUCA or ST, which is much
simpler than previous iterative methods of weight
determinations, and then perform a long MUCA or ST
production run.
The first example is
the {\it replica-exchange multicanonical algorithm} (REMUCA)
\cite{SO3}.
In REMUCA,
a short replica-exchange simulation is performed, and the multicanonical
weight factor is determined by
WHAM \cite{FS2,WHAM}.
Another example of such a combination is the
{\it replica-exchange simulated tempering} (REST) \cite{MO4}.
In REST, a short replica-exchange simulation is performed, and
the simulated tempering weight factor is determined by
WHAM \cite{FS2,WHAM}.
We have introduced a further extension of REMUCA,
 which we refer to as {\it multicanonical replica-exchange method}
 (MUCAREM) \cite{SO3}.
In MUCAREM, the multicanonical weight factor is first
determined as in REMUCA, and then
 a replica-exchange multicanonical production simulation is performed
 with a small number of replicas.

In this article, we first describe the multidimensional
replica-exchange method, a particular realization of
which is the replica-exchange umbrella 
sampling \cite{SKO}.  We then present the
replica-exchange multicanonical algorithm \cite{SO3}.
The effectiveness of these methods is tested with short
peptide systems.

\section{Methods}

\subsection{Multidimensional Replica-Exchange Method}
We first review the original 
{\it replica-exchange method}
(REM) \cite{RE1,RE2} (see Ref.~\cite{SO} for 
details of the molecular dynamics version). 

We consider a system of $N$ atoms 
with their coordinate vectors and
momentum vectors denoted by 
$q \equiv \{{\biq}_1, \cdots, {\biq}_N\}$ and 
$p \equiv \{{\bip}_1, \cdots, {\bip}_N\}$,
respectively.
The Hamiltonian $H(q,p)$ of the system is the sum of the
kinetic energy $K(p)$ and the potential energy $E(q)$:
\begin{equation}
H(q,p) =  K(p) + E(q)~.
\label{eqn1}
\end{equation}
In the canonical ensemble at temperature $T$ 
each state $x \equiv (q,p)$ with the Hamiltonian $H(q,p)$
is weighted by the Boltzmann factor:
\begin{equation}
W_T(x) = e^{-\beta H(q,p)}~,
\label{eqn3}
\end{equation}
where the inverse temperature $\beta$ is defined by 
$\beta = 1/k_B T$ ($k_B$ is the Boltzmann constant). 

The generalized ensemble for REM consists of 
$M$ {\it non-interacting} copies (or, replicas) 
of the original system in the canonical ensemble
at $M$ different temperatures $T_m$ ($m=1, \cdots, M$).
We arrange the replicas so that there is always
exactly one replica at each temperature.
Then there is  a one-to-one correspondence between replicas
and temperatures; the label $i$ ($i=1, \cdots, M$) for replicas 
is a permutation of 
the label $m$ ($m=1, \cdots, M$) for temperatures,
and vice versa:
\begin{equation}
\left\{
\begin{array}{rl}
i &=~ i(m) ~\equiv~ f(m)~, \cr
m &=~ m(i) ~\equiv~ f^{-1}(i)~,
\end{array}
\right.
\label{eqn4b}
\end{equation}
where $f(m)$ is a permutation function of $m$ and
$f^{-1}(i)$ is its inverse.

Let $X = \left\{x_1^{[i(1)]}, \cdots, x_M^{[i(M)]}\right\} 
= \left\{x_{m(1)}^{[1]}, \cdots, x_{m(M)}^{[M]}\right\}$ 
stand for a ``state'' in this generalized ensemble.
Here, the superscript and the subscript in $x_m^{[i]}$
label the replica and the temperature, respectively.
The state $X$ is specified by the $M$ sets of 
coordinates $q^{[i]}$ and momenta $p^{[i]}$
of $N$ atoms in replica $i$ at temperature $T_m$:
\begin{equation}
x_m^{[i]} \equiv \left(q^{[i]},p^{[i]}\right)_m~.
\label{eqn5}
\end{equation}
Because the replicas are non-interacting, the weight factor for
the state $X$ 
is given by
the product of Boltzmann factors for each replica (or at each
temperature):
\begin{equation}
\begin{array}{rl}
W_{\rm REM}(X) &= \exp \left\{- \dis{\sum_{i=1}^M \beta_{m(i)} 
H\left(q^{[i]},p^{[i]}\right) } \right\}~, \cr
 &= \exp \left\{- \dis{\sum_{m=1}^M \beta_m 
H\left(q^{[i(m)]},p^{[i(m)]}\right) }
 \right\}~,
\end{array}
\label{eqn7}
\end{equation}
where $i(m)$ and $m(i)$ are the permutation functions in 
Eq.~(\ref{eqn4b}).

We now consider exchanging a pair of replicas.
Suppose we exchange replicas $i$ and $j$ which are
at temperatures $T_m$ and $T_n$, respectively:  
\begin{equation}
X = \left\{\cdots, x_m^{[i]}, \cdots, x_n^{[j]}, \cdots \right\} 
\longrightarrow \ 
X^{\prime} = \left\{\cdots, x_m^{[j] \prime}, \cdots, x_n^{[i] \prime}, 
\cdots \right\}~. 
\label{eqn8}
\end{equation}
The exchange of replicas can be written in more detail as
\begin{equation}
\left\{
\begin{array}{rl}
x_m^{[i]} \equiv \left(q^{[i]},p^{[i]}\right)_m & \longrightarrow \ 
x_m^{[j] \prime} \equiv \left(q^{[j]},p^{[j] \prime}\right)_m~, \cr
x_n^{[j]} \equiv \left(q^{[j]},p^{[j]}\right)_n & \longrightarrow \ 
x_n^{[i] \prime} \equiv \left(q^{[i]},p^{[i] \prime}\right)_n~,
\end{array}
\right.
\label{eqn9}
\end{equation}
where the momenta are uniformly rescaled according to \cite{SO}
\begin{equation}
\left\{
\begin{array}{rl}
p^{[i] \prime} & \equiv \dis{\sqrt{\frac{T_n}{T_m}}} ~p^{[i]}~, \cr
p^{[j] \prime} & \equiv \dis{\sqrt{\frac{T_m}{T_n}}} ~p^{[j]}~.
\end{array}
\right.
\label{eqn11}
\end{equation}

In order for this exchange process to converge towards the equilibrium
distribution based on Eq.~(\ref{eqn7}), it is sufficient to 
impose the detailed balance
condition on the transition probability $w(X \rightarrow X^{\prime})$:
\begin{equation}
W_{\rm REM}(X) \  w(X \rightarrow X^{\prime})
= W_{\rm REM}(X^{\prime}) \  w(X^{\prime} \rightarrow X)~.
\label{eqn12}
\end{equation}
From Eqs.~(\ref{eqn1}), (\ref{eqn7}), (\ref{eqn11}), 
and (\ref{eqn12}), we have
\begin{equation}
\dis{\frac{w(X \rightarrow X^{\prime})} 
     {w(X^{\prime} \rightarrow X)}} 
= \exp \left( - \Delta \right)~,
\label{eqn13}
\end{equation}
where
\begin{eqnarray}
\Delta &=& \beta_m 
\left(E\left(q^{[j]}\right) - E\left(q^{[i]}\right)\right) 
- \beta_n
\left(E\left(q^{[j]}\right) - E\left(q^{[i]}\right)\right)~,
\label{eqn14a} \\
  &=& \left(\beta_m - \beta_n \right)
\left(E\left(q^{[j]}\right) - E\left(q^{[i]}\right)\right)~. 
\label{eqn14b}
\end{eqnarray}
This can be satisfied, for instance, by the usual Metropolis criterion:
\begin{equation}
w(X \rightarrow X^{\prime}) \equiv
w\left( x_m^{[i]} ~\left|~ x_n^{[j]} \right. \right) 
= \left\{
\begin{array}{ll}
 1~, & {\rm for} \ \Delta \le 0~, \cr
 \exp \left( - \Delta \right)~, & {\rm for} \ \Delta > 0~.
\end{array}
\right.
\label{eqn15}
\end{equation}
Note that because of the velocity rescaling of Eq.~(\ref{eqn11})
the kinetic energy terms are cancelled out in Eqs.~(\ref{eqn14a})
(and (\ref{eqn14b})) and that 
the same criterion, Eqs.~(\ref{eqn14b}) and (\ref{eqn15}),
which was originally
derived for Monte Carlo algorithm \cite{RE1,RE2}
is recovered \cite{SO}.

A simulation of the 
{\it replica-exchange method} (REM) \cite{RE1,RE2}
is then realized by alternately performing the following two
steps:
\begin{enumerate}
\item Each replica in canonical ensemble of the fixed temperature 
is simulated $simultaneously$ and $independently$
for a certain MC or MD steps. 
\item A pair of replicas,
say $x_m^{[i]}$ and $x_{n}^{[j]}$, are exchanged
with the probability
$w\left( x_m^{[i]} ~\left|~ x_{n}^{[j]} \right. \right)$ 
in Eq.~(\ref{eqn15}).
\end{enumerate}
In the present work, we employ molecular dynamics algorithm
for Step 1.
Note that in Step 2 we exchange only pairs of replicas corresponding to
neighboring temperatures, because
the acceptance ratio of the exchange decreases exponentially
with the difference of the two $\beta$'s (see Eqs.~(\ref{eqn14b})
and (\ref{eqn15})).
Note also that whenever a replica exchange is accepted
in Step 2, the permutation functions in Eq.~(\ref{eqn4b})
are updated.

The method is particularly suitable for parallel
computers.  Because one can minimize the amount of information
exchanged among nodes, it is best to assign each replica to
each node (exchanging pairs of temperature values among
nodes is much faster than exchanging coordinates and
momenta).
This means that we keep track of the permutation function
$m(i;t)=f^{-1}(i;t)$ in Eq.~(\ref{eqn4b}) as a function
of MD step $t$ throughout the simulation.

The major advantage of REM over other generalized-ensemble
methods such as multicanonical algorithm \cite{MUCA,MUCArev}
and simulated tempering \cite{ST1,ST2}
lies in the fact that the weight factor 
is {\it a priori} known (see Eq.~(\ref{eqn7})), whereas
in the latter algorithms the determination of the
weight factors can be very tedius and time-consuming.
A random walk in ``temperature space'' is
realized for each replica, which in turn induces a random
walk in potential energy space.  This alleviates the problem
of getting trapped in states of energy local minima.
In REM, however, the number of required replicas increases
as the system size $N$ increases (according to $\sqrt N$) \cite{RE1}.
This demands a lot of computer power for complex systems.

We now present our multidimensional extension of REM, which
we refer to as {\it multidimensional replica-exchange method}
(MREM).
The crucial observation that led to the new algorithm is:  
As long as we have $M$ {\it non-interacting}
replicas of the original system, the Hamiltonian 
$H(q,p)$ of the system does not have to be identical
among the replicas and it can depend on a parameter
with different parameter values for different replicas.
Namely, we can write the Hamiltonian for the $i$-th
replica at temperature $T_m$ as
\begin{equation}
H_m (q^{[i]},p^{[i]}) =  K(p^{[i]}) + E_{\lambda_m} (q^{[i]})~,
\label{Eqn16}
\end{equation}
where the potential energy $E_{\lambda_m}$ depends on a
parameter $\lambda_m$ and can be written as
\begin{equation}
E_{\lambda_m} (q^{[i]}) = E_0 (q^{[i]}) + \lambda_m V(q^{[i]})~.
\label{Eqn16p}
\end{equation}
This expression for the potential energy is often used in
simulations.
For instance, in umbrella sampling \cite{US}, $E_0(q)$ and
$V(q)$ can be respectively taken as the original potential
energy and the ``biasing'' potential energy with the
coupling parameter $\lambda_m$.  In simulations of spin 
systems, on the other hand, 
$E_0(q)$ and $V(q)$ (here, $q$ stands for spins)
can be respectively considered as the
zero-field term and the magnetization term coupled with
the external field $\lambda_m$. 

While replica $i$ and temperature
$T_m$ are in one-to-one correspondence
in the original REM,
replica $i$ and ``parameter set''
$\Lambda_m \equiv (T_m,\lambda_m)$ are in one-to-one
correspondence in the new algorithm.
Hence, the present algorithm can be considered as a
multidimensional extension of the original replica-exchange
method where the ``parameter space'' is one-dimensional 
(i.e., $\Lambda_m = T_m$).
Because the replicas are non-interacting, the weight factor 
for the state $X$ in this new
generalized ensemble is again given by the product
of Boltzmann factors for each replica (see Eq.~(\ref{eqn7})):
\begin{equation}
\begin{array}{rl}
W_{\rm MREM}(X) &= \exp \left\{- \dis{\sum_{i=1}^M \beta_{m(i)} 
H_{m(i)}\left(q^{[i]},p^{[i]}\right) } \right\}~, \cr
 &= \exp \left\{- \dis{\sum_{m=1}^M \beta_m 
H_m\left(q^{[i(m)]},p^{[i(m)]}\right) }
 \right\}~,
\end{array}
\label{Eqn19}
\end{equation}
where $i(m)$ and $m(i)$ are the permutation functions in 
Eq.~(\ref{eqn4b}).
Then the same derivation
that led to the original replica-exchange
criterion follows, and the
transition probability of replica exchange is
given by Eq.~(\ref{eqn15}), where
we now have (see Eq.~(\ref{eqn14a}))
\begin{equation}
\Delta = \beta_m 
\left(E_{\lambda_m}\left(q^{[j]}\right) - 
E_{\lambda_m}\left(q^{[i]}\right)\right) 
- \beta_n
\left(E_{\lambda_n}\left(q^{[j]}\right) - 
E_{\lambda_n}\left(q^{[i]}\right)\right)~.
\label{Eqn21}
\end{equation}
Here, $E_{\lambda_m}$ and $E_{\lambda_n}$ are the
total potential energies (see Eq.~(\ref{Eqn16p})).
Note that we need to newly evaluate the potential
energy for exchanged coordinates,
$E_{\lambda_m} (q^{[j]})$ and $E_{\lambda_n} (q^{[i]})$,
because $E_{\lambda_m}$ and $E_{\lambda_n}$ are in general
different functions.

For obtaining the canonical distributions,
the weighted histogram analysis method (WHAM) \cite{FS2,WHAM}
is particularly suitable.
Suppose we have made a single run of the present
replica-exchange simulation with $M$ replicas that correspond
to $M$ different parameter sets
$\Lambda_m \equiv (T_m,\lambda_m)$ ($m=1, \cdots, M$).
Let $N_m(E_0,V)$ and $n_m$
be respectively 
the potential-energy histogram and the total number of
samples obtained for the $m$-th parameter set
$\Lambda_m$.
The WHAM equations that yield the canonical
probability distribution $P_{T,\lambda} (E_0,V)$
with any
potential-energy parameter value $\lambda$ at
any temperature $T=1/k_B \beta$
are then given by \cite{FS2,WHAM}
\begin{equation}
P_{T,\lambda} (E_0,V)
= \left[\frac{\dis{\sum_{m=1}^M g_m^{-1}~N_m(E_0,V)}} 
{\dis{\sum_{m=1}^M g_m^{-1}~n_{m}~e^{f_m-\beta_m E_{\lambda_m}}}} \right]
e^{-\beta E_{\lambda}}~,
\label{eqn19}
\end{equation}
and
\begin{equation}
e^{-f_m} = \sum_{E_0,V} P_{T_m,\lambda_m} (E_0,V)~.
\label{eqn20}
\end{equation}
Here, 
$g_m = 1 + 2 \tau_m$, 
and $\tau_m$ is the integrated
autocorrelation time at temperature $T_m$ with
the parameter value $\lambda_m$.
Note that the unnormalized probability distribution
$P_{T,\lambda}(E_0,V)$ and the ``dimensionless''
Helmholtz free energy $f_m$ in Eqs.~(\ref{eqn19}) and
(\ref{eqn20}) are solved self-consistently
by iteration \cite{FS2,WHAM}.  

We can use this new replica-exchange method for 
free energy calculations.  We first describe the free-energy
perturbation case.  The potential energy is given by
\begin{equation}
E_{\lambda} (q) = E_I (q) + \lambda \left(E_F (q) - E_I (q)\right)~,
\label{Eqn22}
\end{equation}
where $E_I$ and $E_F$ are the potential energy for
a ``wild-type'' molecule and a ``mutated''
molecule, respectively.  Note that this equation has the same
form as Eq.~(\ref{Eqn16p}).

Our replica-exchange simulation is performed for $M$ replicas
with $M$ different values of the parameters
$\Lambda_m = (T_m,\lambda_m)$.
Since $E_{\lambda = 0} (q) = E_I (q)$ and
$E_{\lambda = 1} (q) = E_F (q)$, we should choose enough
$\lambda_m$ values distributed in the range between 0 and 1
so that we may have sufficient replica exchanges.
From the simulation, $M$ histograms $N_m (E_I,E_F-E_I)$, or
equivalently $N_m(E_I,E_F)$, are obtained.  The Helmholtz
free energy difference of ``mutation'' at 
temperature $T$,
$\Delta F \equiv F_{\lambda = 1} - F_{\lambda = 0}$, can then
be calculated from 
\begin{equation}
\exp (-\beta \Delta F) = \frac{Z_{T,\lambda=1}} 
{Z_{T,\lambda=0}} =  
\frac{\dis{\sum_{E_I,E_F}
P_{T,\lambda=1} (E_I,E_F)}}
{\dis{\sum_{E_I,E_F}
P_{T,\lambda=0} (E_I,E_F)}} ~,
\label{Eqn24}
\end{equation}
where $P_{T,\lambda} (E_I,E_F)$ are obtained from the WHAM
equations of Eqs.~(\ref{eqn19}) and (\ref{eqn20}).

We now describe another free energy calculations based on
MREM applied to umbrella sampling \cite{US},
which we refer to as 
{\it replica-exchange umbrella sampling} (REUS).
The potential energy is a generalization of Eq.~(\ref{Eqn16p})
and is given by
\begin{equation}
E_{\bil} (q) = E_0 (q) + \sum_{\ell = 1}^L
\lambda^{(\ell)} V_{\ell} (q)~,
\label{Eqn25}
\end{equation}
where $E_0(q)$ is the original unbiased potential, 
$V_{\ell}(q)$ ($\ell =1, \cdots, L$) are the 
biasing (umbrella) potentials, and $\lambda^{(\ell)}$ are the
corresponding coupling constants
($\bil = (\lambda^{(1)}, \cdots, \lambda^{(L)})$).
Introducing a ``reaction coordinate'' $\xi$,
the umbrella potentials are usually written as harmonic
restraints:
\begin{equation}
V_{\ell} (q) = k_{\ell} \left[ \xi (q) - d_{\ell} \right]^2~,
~(\ell =1, \cdots, L)~,
\label{Eqn26}
\end{equation}
where $d_{\ell}$ are the midpoints and $k_{\ell}$ are the
strengths of the restraining potentials.
We prepare $M$ replicas with $M$
different values of the parameters
$\biL_m = (T_m,\bil_m)$, and the replica-exchange
simulation is performed.  Since the umbrella potentials
$V_{\ell} (q)$ in Eq.~(\ref{Eqn26})
are all functions of the reaction coordinate
$\xi$ only, we can take the histogram
$N_m (E_0,\xi)$ instead of
$N_m (E_0,V_1, \cdots, V_L)$.
The WHAM equations of
Eqs.~(\ref{eqn19}) and (\ref{eqn20}) can then be written as
\begin{equation}
P_{T,\bil} (E_0,\xi)
= \left[\frac{\dis{\sum_{m=1}^M g_m^{-1}~N_m(E_0,\xi)}} 
{\dis{\sum_{m=1}^M g_m^{-1}~n_{m}~e^{f_m-\beta_m E_{\bil_m}}}} \right]
e^{-\beta E_{\bil}}~,
\label{Eqn27}
\end{equation}
and
\begin{equation}
e^{-f_m} = \sum_{E_0,\xi} P_{T_m,\bil_m} (E_0,\xi)~.
\label{Eqn28}
\end{equation}
The expectation value of a physical quantity $A$ is now
given by
\begin{equation}
<A>_{T,\bil} \ = \frac{\dis{\sum_{E_0,\xi}
A(E_0,\xi) P_{T,\bil} (E_0,\xi)}}
{\dis{\sum_{E_0,\xi} P_{T,\bil} (E_0,\xi)}}~.
\label{Eqn29}
\end{equation}

The potential of mean force (PMF), or free energy as a function of
the reaction coordinate, of the original, unbiased system 
at temperature $T$ is given by
\begin{equation}
{\cal W}_{T,\bil = \{0\}} (\xi) = - k_B T \ln
\left[ \sum_{E_0} P_{T,\bil = \{0\}} (E_0,\xi) \right]~,
\label{Eqn30}
\end{equation}
where $\{0\} = (0, \cdots, 0)$.

\subsection{Replica-Exchange Multicanonical Algorithm}
We first briefly review the multicanonical
algorithm \cite{MUCA,MUCArev}.
Because the coordinates $q$ and momenta $p$ are decoupled
in Eq.~(\ref{eqn1}), we can suppress the kinetic energy
part and can write the Boltzmann factor as
\begin{equation}
W_T(x) = W_T(E) = e^{-\beta E}~.
\label{EQ4b}
\end{equation}
The canonical probability distribution of potential energy
$P_T(E)$ is then given by the product of the
density of states $n(E)$ and the Boltzmann weight factor
$W_T(E)$:
\begin{equation}
 P_T(E) \propto n(E) W_T(E)~.
\label{EQ4c}
\end{equation}

In the multicanonical ensemble (MUCA) \cite{MUCA,MUCArev}, on the other
hand, each state is weighted by
a non-Boltzmann weight
factor $W_{\rm mu}(E)$ (which we refer to as the {\it multicanonical
weight factor}) so that a uniform energy
distribution $P_{\rm mu}(E)$ is obtained:
\begin{equation}
 P_{\rm mu}(E) \propto n(E) W_{\rm mu}(E) \equiv {\rm constant}~.
\label{EQ5}
\end{equation}
The flat distribution implies that
a free random walk in the potential energy space is realized
in this ensemble.
This allows the simulation to escape from any local minimum-energy states
and to sample the configurational space much more widely than the conventional
canonical MC or MD methods.

From the definition in Eq.~(\ref{EQ5}) the multicanonical
weight factor is inversely proportional to the density of
states, and we can write it as follows:
\begin{equation}
 W_{\rm mu}(E) \equiv e^{-\beta_0 E_{\rm mu}(E;T_0)}
= \frac{1}{n(E)}~,
\label{EQ6}
\end{equation}
where we have chosen an arbitrary reference
temperature, $T_0 = 1/k_{\rm B} \beta_0$, and
the ``{\it multicanonical potential energy}''
is defined by
\begin{equation}
 E_{\rm mu}(E;T_0) = k_{\rm B} T_0 \ln n(E) = T_0 S(E)~.
\label{EQ7}
\end{equation}
Here, $S(E)$ is the entropy in the microcanonical
ensemble.

A multicanonical Monte Carlo simulation
is performed, for instance, with the usual Metropolis criterion:
The transition probability of state $x$ with potential energy
$E$ to state $x^{\prime}$ with potential energy $E^{\prime}$ is given by
\begin{equation}
w(x \rightarrow x^{\prime}) = \left\{
\begin{array}{ll}
 1~, & {\rm for} \ \Delta E_{\rm mu} \le 0~, \cr
 \exp \left( - \beta_0 \Delta E_{\rm mu} \right)~, & {\rm for} \
\Delta E_{\rm mu} > 0~,
\end{array}
\right.
\label{EQ8}
\end{equation}
where
\begin{equation}
\Delta E_{\rm mu} \equiv E_{\rm mu}(E^{\prime};T_0) - E_{\rm mu}(E;T_0)~.
\label{EQ9}
\end{equation}
The molecular dynamics algorithm in multicanonical ensemble
also naturally follows from Eq.~(\ref{EQ6}), in which the
regular constant temperature molecular dynamics simulation
(with $T=T_0$) is performed by solving the following modified
Newton equation: \cite{HOE96,NNK,BK} 
\begin{equation}
\dot{{\bip}}_k ~=~ - \frac{\partial E_{\rm mu}(E;T_0)}{\partial {\biq}_k}
~=~ \frac{\partial E_{\rm mu}(E;T_0)}{\partial E}~{\bif}_k~,
\label{EQ9a}
\end{equation}
where ${\bif}_k$ is the usual force acting on the $k$-th atom
($k = 1, \cdots, N$).
From Eq.~(\ref{EQ7}) this equation can be rewritten as
\begin{equation}
\dot{{\bip}}_k ~=~ \frac{T_0}{T(E)}~{\bif}_k~,
\label{EQ9b}
\end{equation}
where the following thermodynamic relation gives
the definition of the ``effective temperature''
$T(E)$:
\begin{equation}
\left. \frac{\partial S(E)}{\partial E}\right|_{E=E_a}
~=~ \frac{1}{T(E_a)}~,
\label{EQ9c}
\end{equation}
with
\begin{equation}
E_a ~=~ <E>_{T(E_a)}~.
\label{EQ9d}
\end{equation}

The multicanonical weight factor is usually determined by
iterations of short trial simulations.  The details of this process
are described, for instance, in Refs.~\cite{MUCArev,OH}.
However, the iterative process can be non-trivial and very tedius for
complex systems.

After the optimal multicanonical weight factor is determined, one
performs a long multicanonical simulation once.  By monitoring
the potential energy throughout the simulation, one can find the
global-minimum-energy state.
Moreover, by using the obtained histogram $N_{\rm mu}(E)$ of the
potential energy distribution $P_{\rm mu}(E)$, 
the expectation value of a physical quantity $A$
at any temperature $T=1/k_{\rm B} \beta$
can be calculated from
\begin{equation}
<A>_T \ = \frac{\dis{\sum_E A(E)~n(E)~e^{-\beta E}}}
{\dis{\sum_E n(E)~e^{-\beta E}}}~,
\label{EQ18}
\end{equation}
where the best estimate of the density of states
$n(E)$ is given by 
the single-histogram
reweighting techniques (see Eq.~(\ref{EQ5})) \cite{FS1}:
\begin{equation}
 n(E) = \frac{N_{\rm mu}(E)}{W_{\rm mu}(E)}~.
\label{EQ17}
\end{equation}

The {\it replica-exchange multicanonical algorithm} (REMUCA) 
\cite{SO3} overcomes
both the difficulties of MUCA (the multicanonical weight factor
determination is non-trivial)
and REM (a lot of replicas, or computation time, is required).
In REMUCA we first perform a short REM simulation (with $M$ replicas)
to determine the
multicanonical weight factor and then perform with this weight
factor a regular multicanonical simulation with high statistics.
The first step is accomplished by the weighted histogram 
analysis method \cite{FS2,WHAM}.
Let $N_m(E)$ and $n_m$ be respectively
the potential-energy histogram and the total number of
samples obtained at temperature $T_m=1/k_{\rm B} \beta_m$ of the REM run.
The density of states $n(E)$ is then given by solving 
the following WHAM equations
\cite{FS2,WHAM}:
\begin{equation}
n(E) = \frac{\dis{\sum_{m=1}^M ~g_m^{-1}~N_m(E)}}
{\dis{\sum_{m=1}^M ~g_m^{-1}~n_m~e^{f_m-\beta_m E}}}~,
\label{EQ8a}
\end{equation}
where
\begin{equation}
e^{-f_m} = \sum_{E} ~n(E)~e^{-\beta_m E}~.
\label{EQ8b}
\end{equation}

Once the estimate of the density of states is obtained, the
multicanonical weight factor can be directly determined from
Eq.~(\ref{EQ6}) (see also Eq.~(\ref{EQ7})).
Actually, the multicanonical potential energy $E_{\rm mu}(E;T_0)$ 
thus determined is only reliable in the following range:
\begin{equation}
E_1 \le E \le E_M~,
\label{EQ29}
\end{equation}
where 
\begin{equation}
\left\{
\begin{array}{rl}
E_1 &=~ <E>_{T_1}~, \\
E_M &=~ <E>_{T_M}~,
\end{array}
\right.
\label{EQ29b}
\end{equation}
and $T_1$ and $T_M$ are respectively the lowest and the highest
temperatures used in the REM run.
Outside this range we extrapolate
the multicanonical potential energy linearly:
\begin{equation}
 {\cal E}_{\rm mu}^{\{0\}}(E) \equiv \left\{
   \begin{array}{@{\,}ll}
   \left. \dis{\frac{\partial E_{\rm mu}(E;T_0)}{\partial E}}
        \right|_{E=E_1} (E - E_1)
             + E_{\rm mu}(E_1;T_0)~,~~~~~
         {\rm for}~E~&<~E_1~, \\
         E_{\rm mu}(E;T_0)~,~~~~~~~~~~~~~~~
          ~~~~~~~~~~~~~~~~~~~~~~
         {\rm for}~E_1~\le~E~&\le~E_M~, \\
   \left. \dis{\frac{\partial E_{\rm mu}(E;T_0)}{\partial E}}
        \right|_{E=E_M} (E - E_M)
             + E_{\rm mu}(E_M;T_0)~,~
         {\rm for}~E~&>~E_M~.
   \end{array}
   \right.
\label{EQ31}
\end{equation}
A long multicanonical MD run is then performed by solving
the Newton equations in Eq.~(\ref{EQ9a}) 
into which we substitute ${\cal E}_{mu}^{\{0\}}(E)$ of
Eq.~(\ref{EQ31}).
Finally, the results are analyzed by the single-histogram
reweighting techniques as described in Eq.~(\ref{EQ17})
(and Eq.~(\ref{EQ18})).
We remark that
our multicanonical MD simulation here
actually results in a canonical simulation at
$T=T_1$ for $E < E_1$, a multicanonical simulation for
$E_1 \le E \le E_M$, and a canonical simulation at
$T=T_M$ for $E > E_M$ (a detailed discussion on this
point is given in Ref.~\cite{RevMSO}).
Note also that the above arguments are independent of
the value of $T_0$, and we
will get the same results, regardless of its value.

Since the WHAM equations are based on histograms,
the density of states $n(E)$, or the multicanonical 
potential energy 
$E_{\rm mu}(E;T_0)$, will be given 
in discrete
values of the potential energy $E$.
For multicanonical MD simulations, however, we need the
derivative of $E_{\rm mu}(E;T_0)$ with respect to $E$
(see Eq.~(\ref{EQ9a})).
We thus introduce some smooth function to
fit the data.  It is best to fit the derivative
$\frac{\partial E_{\rm mu}(E;T_0)}{\partial E}$
directly rather than $E_{\rm mu}(E;T_0)$ itself.
For this we recall the Newton equation of
Eq.~(\ref{EQ9b}) and the thermodynamic
relation of Eqs.~(\ref{EQ9c}) and (\ref{EQ9d}).
The effective temperature $T(E)$, or the derivative
$\frac{\partial E_{\rm mu}(E;T_0)}{\partial E}$,
can be obtained by fitting the inverse 
of Eq.~(\ref{EQ9d}) by a smooth function,
where $<E>_{T(E)}$ is calculated from
Eq.~(\ref{EQ18}) by solving the WHAM equations
of Eqs.~(\ref{EQ8a}) and (\ref{EQ8b}).
Given its derivative, the multicanonical potential
energy can be obtained by integration:
\begin{equation}
E_{\rm mu}(E;T_0) = 
T_0 \int_{E_1}^{E} \frac{\partial S(E)}{\partial E} dE
= T_0 \int_{E_1}^{E} \frac{dE}{T(E)}~.
\label{EQ10}
\end{equation}
We remark that the same equations were used to obtain
the multicanonical weight factor in Ref.~\cite{H97c},
where $<E>_{T(E)}$ was estimated by simulated
annealing instead of REM.

Finally, although we did not find any difficulty in the case 
of protein systems that we studied,
a single REM run in general may not be able to
give an accurate estimate of the
density of states (like in the case of
a strong first-order phase transition \cite{RE1}).  In such a
case we can still greatly simplify the process of the
multicanonical weight factor determination by
combining the present method with the
previous iterative 
methods \cite{MUCArev,OH}.

The formulation of REMUCA is simple and straightforward, but
the numerical improvement is great, because the weight factor
determination for MUCA becomes very difficult
by the usual iterative processes for complex systems.

\section{RESULTS}
   
We now present some examples of the simulation results by
the algorithms described in the previous section.
Short peptide systems were considered.

For molecular dynamics simulations,
the force-field parameters were taken from the all-atom
versions of AMBER \cite{AMBER1,AMBER2}.
The computer code developed in
Refs. \cite{SK,KHG}, which is based on PRESTO
\cite{PRESTO}, was used. The unit time step was set to 0.5 fs.
The temperature during the canonical MD simulations was 
controlled by the constraint method \cite{HLM,EM}.
Besides gas phase simulations, we have also performed
MD simulations with a distance-dependent dielectric, 
$\epsilon=r$, and
with explicit water molecules of TIP3P model \cite{TIP3P}.

As described in detail in the previous section, in 
generalized-ensemble simulations and subsequent
analyses of the data, potential
energy distributions have to be taken as histograms.
For the bin size of these histograms, we used the values
ranging from 0.5 to 2 kcal/mol, depending on the system
studied.

The first example is a penta peptide, Met-enkephalin,
whose amino-acid sequence is: Tyr-Gly-Gly-Phe-Met.
This peptide in gas phase was studied 
with the force field of AMBER in Ref.~\cite{AMBER1}
by the replica-exchange MD simulation \cite{SO}.
We made an MD
simulation of $2 \times 10^6$ time steps (or, 1.0 ns)
for each replica, starting from an extended conformation. 
We used the following eight temperatures: 
700, 585, 489, 409, 342, 286, 239, and 200 K, 
which are distributed 
exponentially, following the annealing schedule of simulated 
annealing simulations \cite{KONF}.  As is shown below, 
this choice already
gave an optimal temperature distribution.
The replica exchange was tried every 10 fs, and the data 
were stored
just before the replica exchange for later analyses.

As for expectation values of physical quantities at 
various temperatures, we used the weighted histogram 
analysis method of Eqs.~(\ref{EQ8a}) and (\ref{EQ8b}).
We remark that for biomolecular systems the integrated
autocorrelation times $\tau_m$ in the reweighting formulae
(see Eq.~(\ref{EQ8a}))
can safely be set to be a constant \cite{WHAM}, and we
do so throughout the analyses in this section.

In Figure 1 the time series of temperature exchange
(a) and 
the total potential energy (b) for one of
the replicas are shown.
We do observe random walks in both
temperature space and potential energy space.
Note that there is a strong correlation between the behaviors
in Figures 1(a) and 1(b).

\begin{figure}[hbtp]
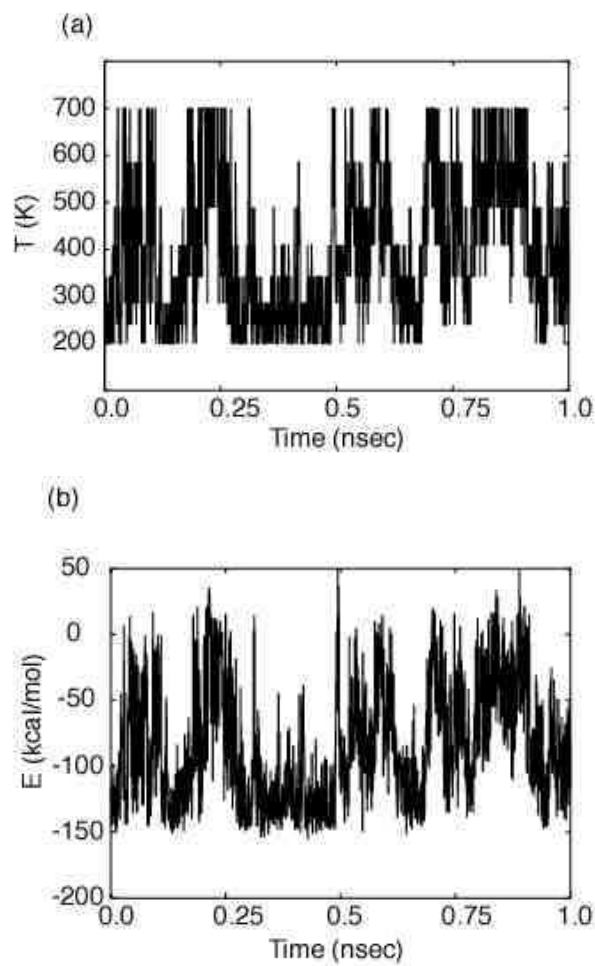

\begin{center}
\includegraphics[width=8.1cm,keepaspectratio]{oka_fig1a.epsf}
\includegraphics[width=8.1cm,keepaspectratio]{oka_fig1b.epsf}
\end{center}
\caption{Time series of (a) temperature exchange
and (b) the total potential energy 
for one of the replicas from a replica-exchange MD simulation
of Met-enkephalin in gas phase.}
\label{fig1}
\end{figure}

In Figure 2 the canonical probability distributions obtained
at the chosen eight temperatures from
the replica-exchange simulation are shown.  We see that there
are enough overlaps between all pairs of distributions, indicating
that there will be sufficient numbers of replica exchanges 
between pairs of replicas.

\begin{figure}[hbtp]
\begin{center}
\includegraphics[width=7.0cm,keepaspectratio]{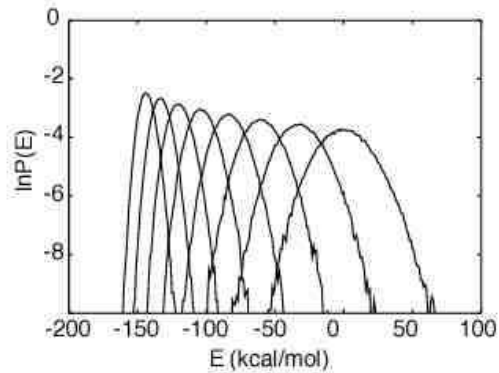}
\end{center}
\caption{The canonical probability distributions of
the total potential energy of Met-enkephalin in gas phase
obtained from
the replica-exchange MD simulation at the eight temperatures.
The distributions
correspond to the following temperatures (from left to right):
200, 239, 286, 342, 409, 489, 585, and 700 K.}
\label{fig2}
\end{figure}

We further compare the results of the replica-exchange simulation
with those of a single canonical MD simulation (of 1 ns)
at the corresponding temperatures.
In Figure 3 we compare the distributions of a pair of 
main-chain dihedral
angles $(\phi,\psi)$ of Gly-2 at two extreme
temperatures ($T=200$ K and 700 K).
While the results at $T=200$ K from the regular canonical simulation
are localized with only one dominant peak, those from the 
replica-exchange simulation have
several peaks (compare Figures 3(a) and 3(b)).
Hence, the replica-exchange
run samples much broader configurational space than the conventional
canonical run at low temperatures.
The results at $T=700$ K (Figures 3(c) and 3(d)), 
on the other hand,
are similar, implying that a regular canonical simulation can
give accurate thermodynamic quantities at high temperatures.

\begin{figure}[hbtp]
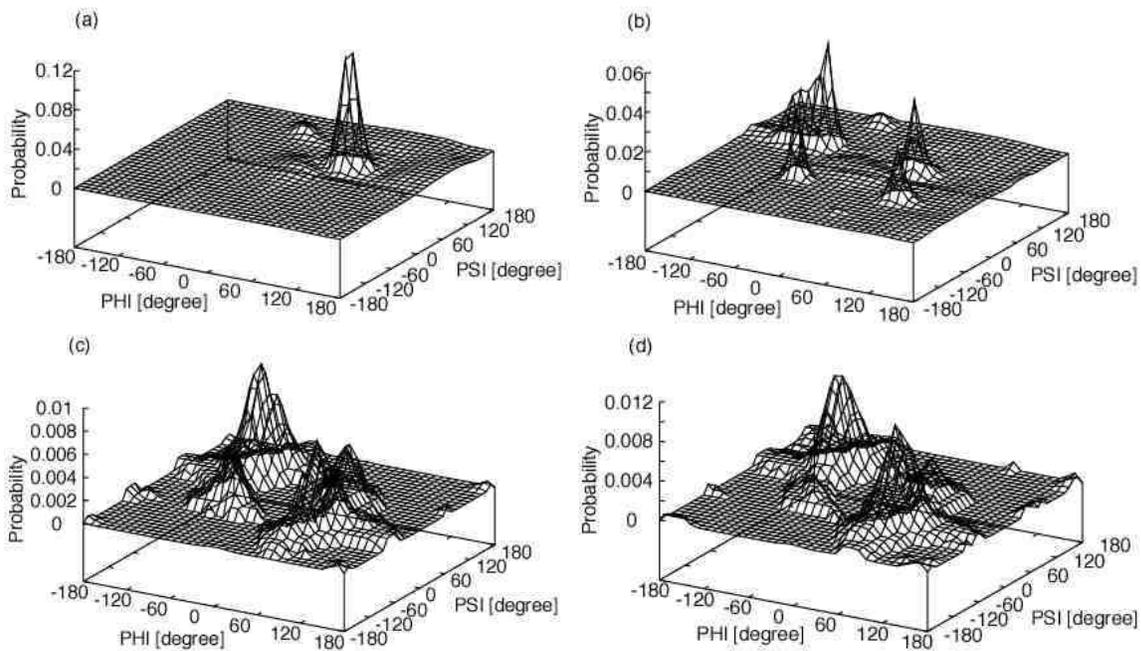

\begin{center}
\includegraphics[width=7.5cm,keepaspectratio]{oka_fig3a.epsf}
\includegraphics[width=7.5cm,keepaspectratio]{oka_fig3b.epsf}
\includegraphics[width=7.5cm,keepaspectratio]{oka_fig3c.epsf}
\includegraphics[width=7.5cm,keepaspectratio]{oka_fig3d.epsf}
\end{center}
\caption{Distributions of a pair of main-chain dihedral
angles $(\phi,\psi)$ of Gly-2 for:
(a) $T=200$ K from a regular canonical MD simulation,
(b) $T=200$ K from the replica-exchange MD simulation,
(c) $T=700$ K from a regular canonical MD simulation, and
(d) $T=700$ K from the replica-exchange MD simulation.}
\label{fig3}
\end{figure}

In Figure 4 we show the average total potential energy 
as a function of temperature.
As expected from the results of Figure 3,
we observe that the canonical simulations at low temperatures
got trapped in states of energy local minima, resulting in
the discrepancies in average values between the results from
the canonical simulations and those from the replica-exchange
simulation.

\begin{figure}[hbtp]
\begin{center}
\includegraphics[width=7.0cm,keepaspectratio]{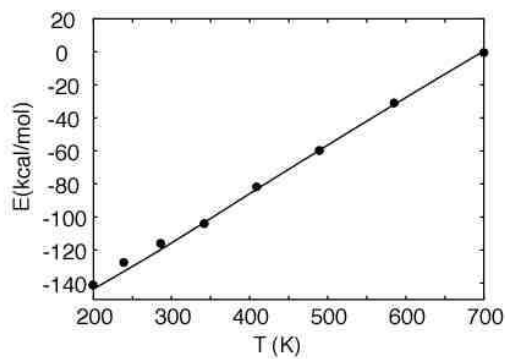}
\end{center}
\caption{Average total potential energy of Met-enkephalin
in gas phase
as a function of temperature.
The solid curve is the result from the replica-exchange
MD simulation and the dots are those of regular canonical
MD simulations.}
\label{fig4}
\end{figure}

We now present the results of
{\it replica-exchange umbrella sampling} (REUS) \cite{SKO}.
The system of a blocked peptide, alanine-trimer, was studied.
Since the thermodynamic
behavior of this peptide was extensively studied by the conventional
umbrella sampling \cite{Boc1}, 
it is a good test case to examine the effectiveness of the new method.
The force field 
parameters were taken from the
all-atom version of AMBER \cite{AMBER1} with a distance-dependent dielectric, 
$\epsilon=r$, which mimics the presence of solvent.  
We made an MD
simulation of $4 \times 10^6$ time steps (or, 2.0 ns)
for each replica, starting from an extended conformation. 
The data were stored every 20 steps (or, 10 fs) for a total of 
$2 \times 10^5$ snapshots. 

In Table 1 we summarize the parameters characterizing the replicas
for the simulations performed. They are one 
original replica-exchange simulation (REM1),
two replica-exchange umbrella sampling simulations (REUS1 and REUS2),
and two conventional umbrella sampling simulations (US1 and US2). 

\begin{table}[hbtp]
\caption{Summary of Parameters in US, REM, and REUS
Simulations}
\begin{center}
\begin{tabular}{lcrlrl} \hline
Run$^a$ & $M^b$ & ${N_T}^b$ & ~~Temperature [K] & $L^b$ & 
~~~~~~~$d_{\ell}$ [\AA] $(k_{\ell}$ [kcal/mol$\cdot$\AA$^2])^c$ \\
  \hline
REM1 & 16 & 16~ & 200, 229, 262, 299, & 0~ &  \\	
       &    &    & 342, 391, 448, 512, &   &   \\
       &    &    & 586, 670, 766, 876, &   &   \\
       &    &    & 1002, 1147, 1311, &   &   \\
       &    &    & 1500 &   &   \\
REUS1, US1 & 14 & 1~ & 300 & 14~ & 0.0 (0.0)$^d$, 1.8 (1.2), 2.8 (1.2), 3.8 (1.2),\\
       &    &   &     &    & 4.8 (1.2), 5.8 (1.2), 6.8 (1.2), 7.8 (1.2),\\
       &    &   &     &    & 8.8 (1.2), 9.8 (1.2), 10.8 (1.2), 11.8 (1.2),\\
       &    &   &     &    & 12.8 (1.2), 13.8 (1.2)\\  
REUS2, US2 & 16 & 4~ & 250, 315, 397, 500 & 4~ & 0.0 (0.0), 7.8 (0.3), 10.8 (0.3), 13.8 (0.3)\\
 \hline
\end{tabular}
\end{center}
\noindent
$^a$ REM, REUS, and US stand for an original replica-exchange
simulation, replica-exchange umbrella sampling simulation, and
conventional umbrella sampling simulation, respectively.
  
\noindent
$^b$ $M$, $N_T$, and $L$ are the total numbers of replicas,
temperatures, and restraining potentials, respectively 
(see Eqs.~(\ref{Eqn19}) and (\ref{Eqn25})).  In REUS2 and
US2 we set
$M=N_T \times L$ for simplicity.  We remark that this
relation is not always required.  For instance,
the 16 replicas could have 16 different temperatures with 16
different restraining potentials (i.e., $M=N_T=L=16$). 

\noindent
$^c$ $d_{\ell}$ and $k_{\ell}$ $(\ell = 1, \cdots, L)$ 
are the strengths and the midpoints
of the restraining potentials, respectively (see Eq.~(\ref{Eqn26})).

\noindent
$^d$ The parameter value 0.0 (0.0) means that the restraining
potential is null, i.e., $V_{\ell}=0$.
\label{Tab1}
\end{table}

The purpose of the present simulations is to test the effectiveness
of the replica-exchange umbrella sampling with respect to the conventional
umbrella sampling (REUS1 and REUS2 versus US1 and US2).
The original replica-exchange simulation without umbrella potentials
(REM1) was also made to set a reference standard for comparison.
For REM1
replica exchange was tried every 20 time steps (or, 10 fs),
as in our previous work \cite{SO}.  
For REUS simulations, on the other hand,
replica exchange was tried every 400 steps (or, 200 fs),
which is less frequent than in REM1.
This is  because we wanted to ensure sufficient
time for system relaxation after $\lambda$-parameter
exchange. 
  
In REM1 there are 16 replicas with 16 different temperatures 
listed in Table 1.
The temperatures are again distributed exponentially. 
After every 10 fs of parallel MD simulations, eight pairs of 
replicas corresponding to neighboring temperatures
were simultaneously exchanged, and the pairing was alternated between
the two possible choices \cite{SO}.
   
For umbrella potentials, the O1 to H5 
hydrogen-bonding distance, or ``end-to-end distance,''
was chosen 
as the reaction coordinate $\xi$ and the harmonic restraining potentials
of $\xi$ in Eq.~(\ref{Eqn26}) were imposed.
The force constants, $k_{\ell}$, and 
the midpoint positions, $d_{\ell}$, are listed in Table 1. 
    
In REUS1 and US1, 14 replicas were simulated with the
same set of umbrella potentials at $T=300$ K. 
Let us order the umbrella potentials, $V_{\ell}$ in Eq.~(\ref{Eqn25}),
in the increasing order of the midpoint value $d_{\ell}$,
i.e., the same order that appears in Table 1.
We prepared replicas so that the potential energy for each
replica includes exactly one umbrella potential
(here, we have $M = L = 14$).
Namely, in Eq.~(\ref{Eqn25}) for $\bil = \bil_m$ we set
\begin{equation}
\lambda^{(\ell)}_m = \delta_{\ell,m}~,
\label{Eqn31}
\end{equation}
where $\delta_{k,l}$ is Kronecker's delta function, and
we have
\begin{equation}
E_{\bil_m} (q^{[i]}) = E_0 (q^{[i]}) + V_m (q^{[i]})~.
\label{Eqn32}
\end{equation}
The difference between REUS1 and US1 is whether replica exchange
is performed or not during the parallel MD simulations.
In REUS1 seven pairs of 
replicas corresponding to ``neighboring'' umbrella potentials,
$V_{m}$ and $V_{m+1}$,
were simultaneously exchanged after every 200 fs of parallel
MD simulations, and the pairing was alternated between
the two possible choices.  (Other pairings will have much smaller 
acceptance ratios of replica exchange.)
The acceptance criterion for replica exchange is given
by Eq.~(\ref{eqn15}), where Eq.~(\ref{Eqn21}) now reads
(with the fixed inverse temperature $\beta = 1/300 k_B$)
\begin{equation}
\Delta = \beta 
\left(V_m\left(q^{[j]}\right) - 
      V_m\left(q^{[i]}\right) -
      V_{m+1}\left(q^{[j]}\right) + 
      V_{m+1}\left(q^{[i]}\right)\right)~,
\label{Eqn33}
\end{equation}
where replica $i$ and $j$ respectively have umbrella potentials
$V_m$ and $V_{m+1}$ before the exchange.

In REUS2 and US2, 16 replicas were simulated at four different
temperatures with four different restraining potentials
(there are $L=4$
umbrella potentials at $N_T=4$ temperatures, making the total
number of replicas $M=N_T \times L=16$; see Table 1).
We can introduce the following labeling for the parameters
characterizing the replicas:
\begin{equation}
\begin{array}{rl}
\biL_m = (T_m,\bil_m) & \longrightarrow
\ \biL_{I,J} = (T_I,\bil_J)~. \cr
(m=1, \cdots, M) & \ \ \ \ \ \ \ \ \ \ (I=1, \cdots, N_T,~J=1, \cdots, L)
\end{array}
\label{Eqn34}
\end{equation}
The potential energy is given by Eq.~(\ref{Eqn32})
with the replacement: $m \rightarrow J$.
Let us again order the umbrella potentials, $V_J$,
and the temperatures, $T_I$, in
the same order that appear in Table 1.
The difference between REUS2 and US2 is 
whether replica exchange
is performed or not during the MD simulations.
In REUS2 we performed the following replica-exchange processes alternately
after every 200 fs of parallel MD simulations:
\begin{enumerate}
\item Exchange pairs of replicas corresponding to neighboring temperatures,
$T_I$ and $T_{I+1}$ 
(i.e., exchange replicas $i$ and $j$ that
respectively correspond to parameters
$\biL_{I,J}$ and $\biL_{I+1,J}$).
(We refer to this process as $T$-exchange.)
\item Exchange pairs of replicas corresponding to 
``neighboring'' umbrella potentials,
$V_J$ and $V_{J+1}$ 
(i.e., exchange replicas $i$ and $j$ that
respectively correspond to parameters
$\biL_{I,J}$ and $\biL_{I,J+1}$).
(We refer to this process as $\lambda$-exchange.)
\end{enumerate}
In each of the above processes, two pairs of replicas were simultaneously
exchanged, and the pairing was further 
alternated between the two possibilities.
The acceptance criterion for these 
replica exchanges is given by 
Eq.~(\ref{eqn15}), where Eq.~(\ref{Eqn21}) now reads
\begin{equation}
\Delta = \left(\beta_{I} - \beta_{I+1} \right)
\left(E_0 \left(q^{[j]}\right) 
    + V_J \left(q^{[j]}\right) 
    - E_0 \left(q^{[i]}\right)
    - V_J \left(q^{[i]}\right)\right)~, 
\label{Eqn35}
\end{equation}
for $T$-exchange, and
\begin{equation}
\Delta = \beta_I 
\left(V_J\left(q^{[j]}\right) - 
      V_J\left(q^{[i]}\right) -
      V_{J+1}\left(q^{[j]}\right) + 
      V_{J+1}\left(q^{[i]}\right)\right)~,
\label{Eqn36}
\end{equation}
for $\lambda$-exchange.
By this procedure, the random walk 
in the reaction coordinate space as well as in temperature
space can be realized.
 
We now give the details of the results obtained.
In order to confirm that our REM simulations performed properly,
we have to examine the time series of various quantites and
observe random walks.  
For instance, the time series of temperature exchange 
for one of the replicas
is shown in Figure~5(a).
The corresponding time series of the reaction coordinate $\xi$,
the distance between atoms O1 and H5, for the same replica is shown 
in Figure~5(b). 

\begin{figure}[hbtp]
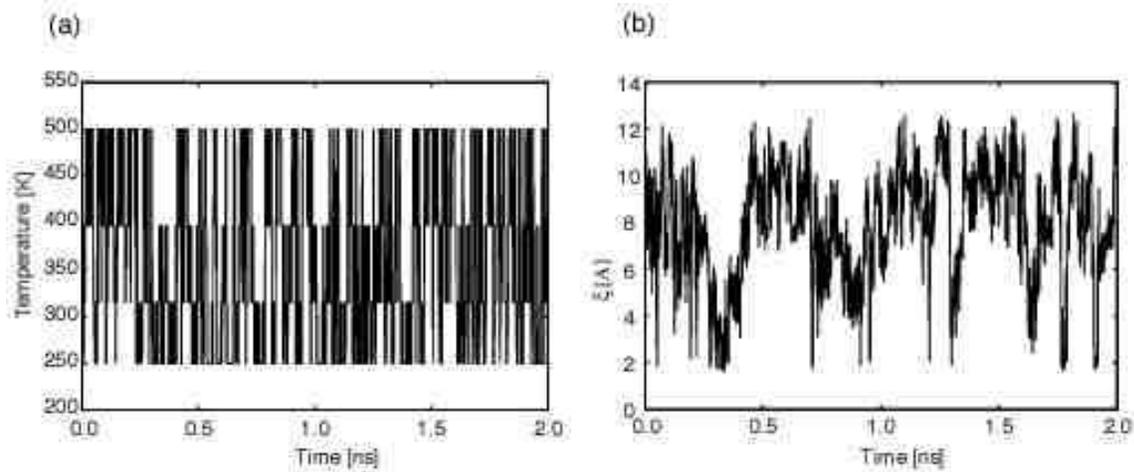

\begin{center}
\includegraphics[width=7.5cm,keepaspectratio]{oka_fig5a.epsf}
\includegraphics[width=7.5cm,keepaspectratio]{oka_fig5b.epsf}
\end{center}
\caption{Time series of (a) temperature exchange for one of the 
replicas 
and (b) the reaction coordinate $\xi$
for the same replica as in (a)
from the replica-exchange umbrella sampling simulation
(REUS2 in Table 1).}
\label{fig5}
\end{figure}

We see that the conformational
sampling along the reaction coordinate is significantly enhanced. 
In the blocked alanine-trimer, the reaction coordinate $\xi$ can be
classified into three regions \cite{Boc1}: 
the helical region ($\xi < 3$ \AA), the turn region
(3 \AA\ $< \xi <$ 7 \AA), and the extended region
($\xi >$ 7 \AA).
Thus, Figure~5(b) implies that helix-coil transitions
frequently occurred during the replica-exchange simulation, whereas
in the conventional canonical 
simulations such a frequent folding and unfolding process cannot be seen.
 
After confirming that the present REUS simulations
performed properly, we now present and compare the
physical quantities
calculated by these simulations.
In Figure~6 the potentials of mean force (PMF) of the
unbiased system along the 
reaction coordinate $\xi$ at $T=300$ K are shown.
The results are from REM1, REUS1, and US1 simulations.
For these calculations, the WHAM equations of Eqs.~(\ref{Eqn27})
and (\ref{Eqn28}) were solved by iteration
first, and then Eq.~(\ref{Eqn30}) was used to obtain the PMF.

\begin{figure}[hbtp]
\begin{center}
\includegraphics[width=7.0cm,keepaspectratio]{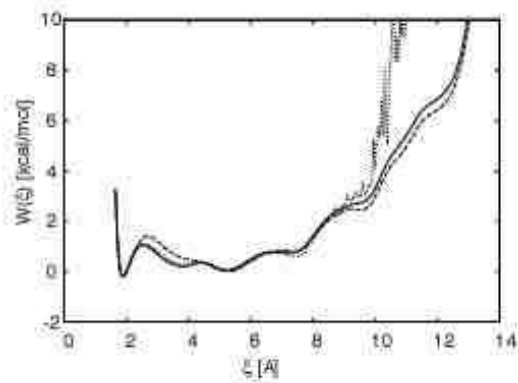}
\end{center}
\caption{The PMF of the unbiased system
along the reaction coordinate $\xi$
at $T=300$ K.
The dotted, solid, and dashed curves were obtained from the
original REM (REM1),
the replica-exchange umbrella sampilng (REUS1), and the conventional
umbrella sampling (US1), respectively.}
\label{fig6}
\end{figure}

From Figure~6 we see that the 
PMF curves obtained by REM1 and REUS1 are 
essentially identical for low values of $\xi$ ($\xi < 7$ \AA).
The two PMF curves start deviating slightly,
as $\xi$ gets larger, and for $\xi > 9$ \AA \ 
the agreement completely deteriorates.
The disagreement comes from the facts that
the average $\xi$ at the highest temperature in REM1 
($T_{16}=1500$ K) 
is $<\xi>_{T_{16}} \approx 8.0$ \AA \ and 
that the original REM
with $T$-exchange only cannot sample accurately the region
where $\xi$ is much larger than 
$<\xi>_{T_{16}}$.
These two simulations were performed under very different
conditions: One was run at different temperatures without
restraining potentials and the other at one temperature
with many restraining potentials (see Table 1).
We thus consider the results to be quite reliable for
($\xi < 9$ \AA).
   
On the other hand, the PMF obtained by US1 is relatively 
larger than those obtained by REM1 and REUS1 in the region 
of 2 \AA\ $< \xi <$ 4 \AA, which corresponds to 
the structural transition state between the $\alpha$-helical 
and turn structures.  This suggests that
US1 got trapped in states of energy local
minima at $T=300$ K. In the region of completely extended structures
($\xi > 9$ \AA), the results of REUS1 and US1 are similar
but the discrepancy is again non-negligible.
We remark that at $T=300$ K the
PMF is the lowest for $\xi \approx 2$ \AA, which
implies that the $\alpha$-helical structure is favored at this 
temperature.
 
We next study
the temperature dependence of physical quantities obtained from the
REM1, REUS2, and US2 simulations.
In Figure~7(a) we show the PMF again at $T=300$ K.  We observe that
the PMF curves from REM1 and REUS2 are essentially identical
for $\xi < 9$ \AA \ and that they deviate
for $\xi > 9$ \AA, because the results for REM1 is not reliable
in this region as noted above.
In fact, by comparing Figures~6 and 7(a), we find that the PMF
obtained from REUS1 and REUS2 are almost in complete agreement
at $T=300$ K in the entire
range of $\xi$ values shown.  On the other hand, we
observe a discrepancy between REUS2 and US2 results.
The PMF curve for US2 is significantly less than that for
REUS2 in the region 2 \AA\ $<\xi<$ 8 \AA.  Note that the 
PMF curves for US1 and US2 are completely in disagreement
(compare Figures~6 and 7(a)).

\begin{figure}[hbtp]
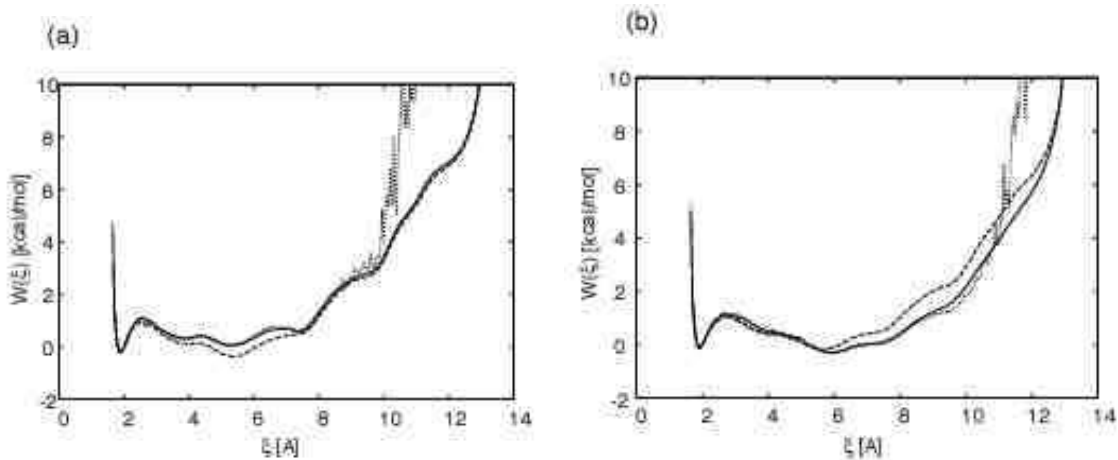

\begin{center}
\includegraphics[width=7.5cm,keepaspectratio]{oka_fig7a.epsf}
\includegraphics[width=7.5cm,keepaspectratio]{oka_fig7b.epsf}
\end{center}
\caption{The PMF of the unbiased system
along the reaction coordinate $\xi$
at two temperatures.
(a) The PMF at $T=300$ K. The dotted, solid, and dashed curves were obtained 
from the original REM (REM1), the replica-exchange umbrella sampling
(REUS2), and the conventional umbrella sampling (US2), respectively.
(b) The PMF at $T=500$ K. The dotted, solid, and dashed curves were obtained 
from the original REM (REM1), the replica-exchange umbrella sampling
(REUS2), and the conventional umbrella sampling (US2), respectively.}
\label{fig7}
\end{figure}

In Figure~7(b) we show the PMF at $T=500$ K, which
we obtained from REM1, REUS2, and US2 simulations.
We again observe that the results from REM1 and REUS2
are in good agreement for a wide range of $\xi$ values.
We find that the results from REM1 do not significantly deteriorate
until $\xi>11$ \AA \ at $T=500$ K, whereas it did start
deviating badly for $\xi >9$ \AA \ at $T=300$ K.  The PMF
curve for US2 deviates strongly from the REUS2 results
for $\xi>6$ \AA \ and is much larger than that of REUS2
(and REM1) in this region.
We remark that at $T=500$ K the
PMF is the lowest for $\xi \approx 6$ \AA \ and
low up to $\xi \approx 8$ \AA,
which implies that extended structures are favored at
this temperature.

In Figure~8 we show the average values of the reaction
coordinate $\xi$ of the unbiased system
as a function of temperature.  The results
are again from the REM1, REUS2, and US2 simulations.
The expectation values were calculated from
Eq.~(\ref{Eqn29}).
We find that the average reaction coordinate, or the
average end-to-end distance, grows as the temperature is raised,
reflecting the unfolding of the peptide upon increased
thermal fluctuations.
Again we observe an agreement between REM1 and REUS2,
whereas the results of US2 deviate.

\begin{figure}[hbtp]
\begin{center}
\includegraphics[width=7.0cm,keepaspectratio]{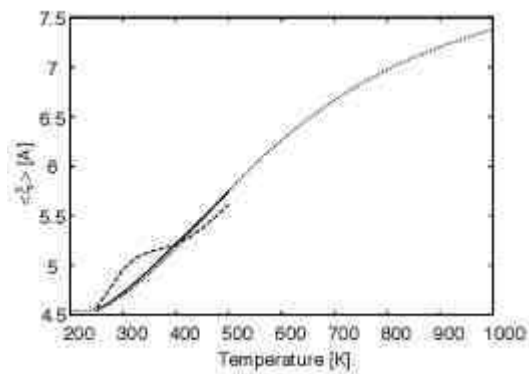}
\end{center}
\caption{Average values of the reaction coordinate $\xi$
of the unbiased system as a function of temperature.
The dotted, solid, and dashed curves were obtained from
the original REM (REM1),
the replica-exchange umbrella sampling (REUS2), and the conventional
umbrella sampling (US2), respectively.
Although the highest temperature in REM1 is 1500 K, only the results
for the temperature range between 200 K and 1000 K are shown for
REM1.
Since the lowest and highest temperatures in REUS2 and US2
are respectively 250 K and 500 K,
only the results between these temperatures are shown for
these simulations.}
\label{fig8}
\end{figure}

We now present the results of another example of the
{\it multidimensional replica-exchange method}.
This time we consider NPT ensemble of argon fluids, and
exchange not only the temperature but also the pressure
values of pairs of replicas during a MC simulation \cite{NSMO}.  
Namely, suppose
we have $M$ replicas with $M$ different values of 
temperature and pressure ($T_m$,$P_m$).  The state $x$ of
replica $i$ is characterized by the scaled coordinates
$\tilde{q}^{[i]}$ and the volume ${\cal V}^{[i]}$ and its weight
is given by
\begin{equation}
W(x) = e^{-\beta_m \left( E(\tilde{q}^{[i]}) + P_m {\cal V}^{[i]}
\right) + N \ln {\cal V}^{[i]}}~.
\label{EQ100}
\end{equation}
The transition probability of replica exchange is then
given by Eq.~(\ref{eqn15}), where
we now have
\begin{equation}
\Delta = \left(\beta_m - \beta_n \right)
\left(E\left(\tilde{q}^{[j]}\right) - E\left(\tilde{q}^{[i]}\right)\right)
+\left(\beta_m P_m - \beta_n P_n \right)
\left({\cal V}^{[j]} - {\cal V}^{[i]}\right)~.
\label{EQ101}
\end{equation}
We prepared $M=64$ replicas with $N_T = 8$ temperature and
$N_P = 8$ pressure values ($M=N_T \times N_P$).  We alternately
exchanged four pairs of temperatures and four pairs of 
pressures during the replica-exchange simulation.
In Figure~9 the values of the set ($T_m$,$P_m$) in reduced units 
are shown as crosses.   

\begin{figure}[hbtp]
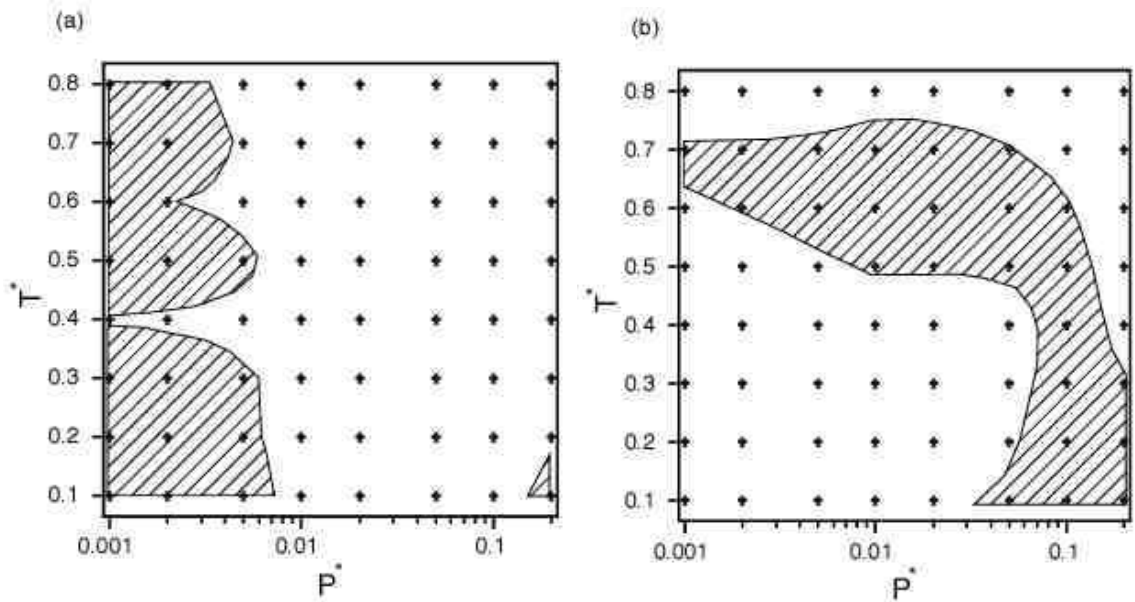

\begin{center}
\includegraphics[width=7.5cm,keepaspectratio]{oka_fig89a.epsf}
\includegraphics[width=7.5cm,keepaspectratio]{oka_fig89b.epsf}
\end{center}
\caption{Regions where the acceptance ratios of replica
exchange become low (shaded regions) for (a) temperature exchange 
and for (b) pressure exchange in the multidimensional
replica-exchange Monte Carlo simulation of argon fluids.
The crosses in the grid indicate the values of the set 
($T_m$,$P_m$) in reduced units.}
\label{fig89}
\end{figure}

The shaded regions in Figure~9 are
where the acceptance ratios of replica exchange become low
($< 20$ \%).  These regions are those where the replica-exchange
method fails due to the existence of first-order phase transitions.
The results of Figure~9 suggest that the multidimensional 
REM enables the simulation to connect regions which cannot
be reached by one-dimensional REM simulations
with only $T$-exchange or
$P$-exchange. 
 
We now present the results of MD simulations based
on {\it replica-exchange multicanonical
algorithm} (REMUCA) \cite{SO3}.
The Met-enkephalin in gas phase was first studied again.
The potential
energy is, however, that of AMBER in Ref.~\cite{AMBER2} instead
of Ref.~\cite{AMBER1}.
In Table 2 we summarize the parameters of the simulations that
were performed.
As discussed in the previous section, REMUCA consists of
two simulations: a short REM simulation (from which the
density of states of the system, or the multicanonical weight factor,
is determined) and a subsequent
production run of MUCA simulation.
The former simulation is referred to as REM1 and the latter
as MUCA1 in Table 2.
Finally, a production run of the original REM simulation
was also performed for comparison and it is referred to as
REM2 in Table 2.

\begin{table}[hbtp]
\caption{Summary of Parameters in REM and REMUCA Simulations}
 \begin{center}
 \begin{tabular}{cccc} \hline
   Run     & No. of Replicas, $M$~ & ~Temperature, $T_m$ (K)
   ($m = 1, \cdots, M$)~ & MD Steps\\
   \hline
   REM1    & 10  & 200, 239, 286, 342, 409, & $2 \times 10^5$ \\
           &     & 489, 585, 700, 836, 1000 & \\
   REM2    & 10  & 200, 239, 286, 342, 409, & $1 \times 10^6$ \\
           &     & 489, 585, 700, 836, 1000 & \\
   MUCA1   & 1   & 1000 & $1 \times 10^7$ \\
   \hline
  \end{tabular}
 \end{center}
 \label{Tab2}
\end{table}

After the simulation of REM1 is finished, we obtained the 
density of states $n(E)$ by the
weighted histogram analysis method of Eqs.~(\ref{EQ8a}) 
and (\ref{EQ8b}).
The density of states will give the average values of the
potential energy from Eq.~(\ref{EQ18}), and we found
\begin{equation}
\left\{
\begin{array}{rl}
E_1 &=~ <E>_{T_1} = -30 ~{\rm kcal/mol}~, \\
E_M &=~ <E>_{T_M} = 195 ~{\rm kcal/mol}~.
\end{array}
\right.
\label{EQ50}
\end{equation}
Then our estimate of the density of states is reliable
in the range $E_1 \le E \le E_M$.
The multicanonical potential energy ${\cal E}_{mu}^{\{0\}}(E)$
was thus determined for the three energy regions
($E < E_1$, $E_1 \le E \le E_M$, and $E > E_M$) from
Eq.~(\ref{EQ31}).
Here, we have set the arbitrary reference temperature to 
be $T_0 = 1000$ K.

After determining the multicanonical weight factor,
we carried out a multicanonical MD simulation of $1 \times
10^7$ steps (or 5 ns) for data collection (MUCA1 in Table 2).
In Figure 10 the probability distribution of
potential energy obtained by 
MUCA1 is plotted.
It can be seen that a good
flat distribution is obtained in the energy region
$E_1 \le E \le E_M$.
In Figure 10 the canonical probability distributions that
were obtained by the reweighting techniques at
$T = T_1 = 200$ K and $T = T_M = 1000$ K are also
shown.
Comparing these curves with those of MUCA1 in
the energy regions $E < E_1$
and $E > E_M$ in Figure 10, we confirm our claim in the
previous section that
MUCA1 gives canonical distributions at $T=T_1$ for
$E < E_1$ and at $T=T_M$ for $E > E_M$, whereas
it gives a multicanonical distribution for
$E_1 \le E \le E_M$.

\begin{figure}[hbtp]
\begin{center}
\includegraphics[width=7.0cm,keepaspectratio]{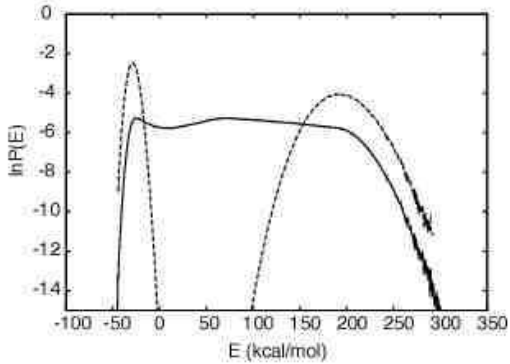}
\end{center}
\caption{Probability distribution of potential energy
        of Met-enkephalin in gas phase that was obtained from 
        the replica-exchange multicanonical simulation (MUCA1
        in Table 2).  The dotted curves are the probability
       distributions of the reweighted
       canonical ensemble at $T = 200$ K (left) and 1000 K (right).}
\label{fig9}
\end{figure}

In the previous
works of multicanonical simulations of Met-enkephalin in gas phase
(see, for instance, Refs.~\cite{HO,MHO}), at least several
iterations of trial simulations were required for the multicanonical
weight determination.
We emphasize that in the present case of REMUCA (REM1), only
one simulation was necessary to determine the optimal multicanonical
weight factor that can cover the energy region corresponding to
temperatures between 200 K and 1000 K.

To check the validity of the canonical-ensemble expectation values
calculated by the new algorithms, we
compare the average potential energy as a function of temperature
in Figure 11.
In REM2 we used the multiple-histogram
techniques (or WHAM) \cite{FS2,WHAM}, whereas the
single-histogram method \cite{FS1}
was used in MUCA1. We can see a perfect coincidence of
these quantities between REM2 and MUCA1 in Figure 11.

\begin{figure}[hbtp]
\begin{center}
\includegraphics[width=7.0cm,keepaspectratio]{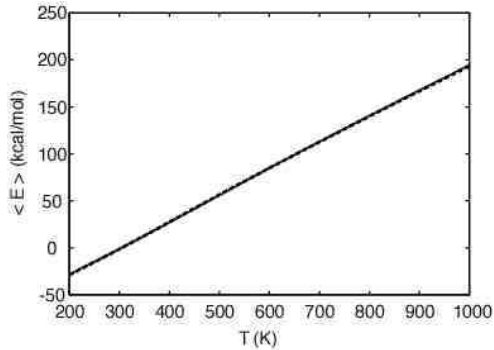}
\end{center}
\caption{The average potential energy of Met-enkephalin in
       gas phase as a function of temperature.
       The solid and dotted curves are obtained from REM2 and
       MUCA1, respectively
       (see Table 2 for the parameters of the simulations).}       
\label{fig10}
\end{figure}

We have so far presented the results of generalized-ensemble
simulations of peptides in gas phase.
However, peptides and proteins are usually in aqueous
solution.  We therefore want to incorporate rigorous
solvation effects in our simulations in order to
compare with experiments.
Met-enkephalin was thus studied by both REM and REMUCA
simulations in aqueous solution based on TIP3P
water model \cite{SO4}.
The AMBER force field of Ref.~\cite{AMBER2} was
used.  The number of water molecules was 526
and they were placed in a sphere of radius of
16 \AA.  Thirty-six replicas that correspond to
temperatures ranging from 200 K to 700 K were
used. 

The time series of the total potential
energy for one of the replicas
is shown in Figure 12.
We do observe a random walk in potential energy space,
which covers an energy range of as much as 2,500 
kcal/mol.

\begin{figure}[hbtp]
\begin{center}
\includegraphics[width=7.0cm,keepaspectratio]{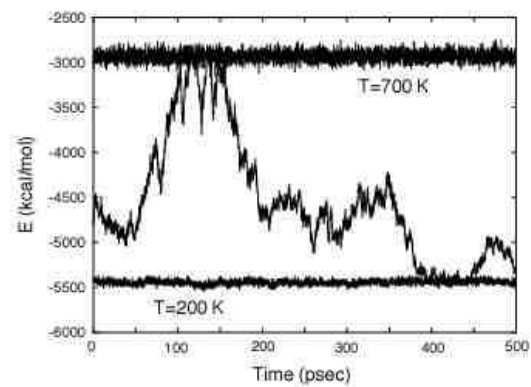}
\end{center}
\caption{Time series of
the total potential energy of Met-enkephalin
in aqueous solution obtained for one of the replicas
from
the replica-exchange MD simulation.  
Corresponding times series in the canonical ensemble
at temperatures 200 K and 700 K are also shown.}
\label{fig11}
\end{figure}

For the REMUCA simulation,
the multicanonical
potential energy and its derivative were obtained by
the weighted histogram analysis method from the
results of a short REM simulation (of 100 psec).
In Figure 13 the probability distribution obtained by 
the multicanonical production run of this REMUCA
simulatoin is plotted.
It can be seen that a good
flat distribution is obtained in the wide
energy range.

\begin{figure}[hbtp]
\begin{center}
\includegraphics[width=7.0cm,keepaspectratio]{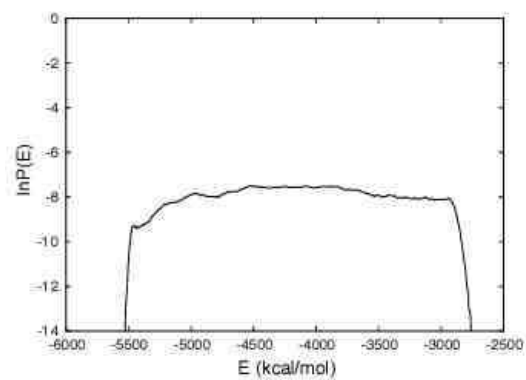}
\end{center}
\caption{Probability distribution of potential energy
        of Met-enkephalin in aqueous solution that was obtained from 
        the replica-exchange multicanonical simulation.}
\label{fig12}
\end{figure}

Finally, in Figure 14 we compare the distributions of a pair of 
main-chain dihedral
angles $(\phi,\psi)$ of Gly-2 and Phe-4 around $T=300$ K between
gas-phase and in-solution results.
While the results in gas phase are well localized
and sharp, those in aqueous solution are
distributed more broadly.  This suggests that the energy
landscape in gas phase is much more rugged than in 
aqueous solution; water considerably smoothes out 
the landscape.  We remark that a similar observation was
made earlier in Ref.~\cite{KOH}. 

\begin{figure}[hbtp]
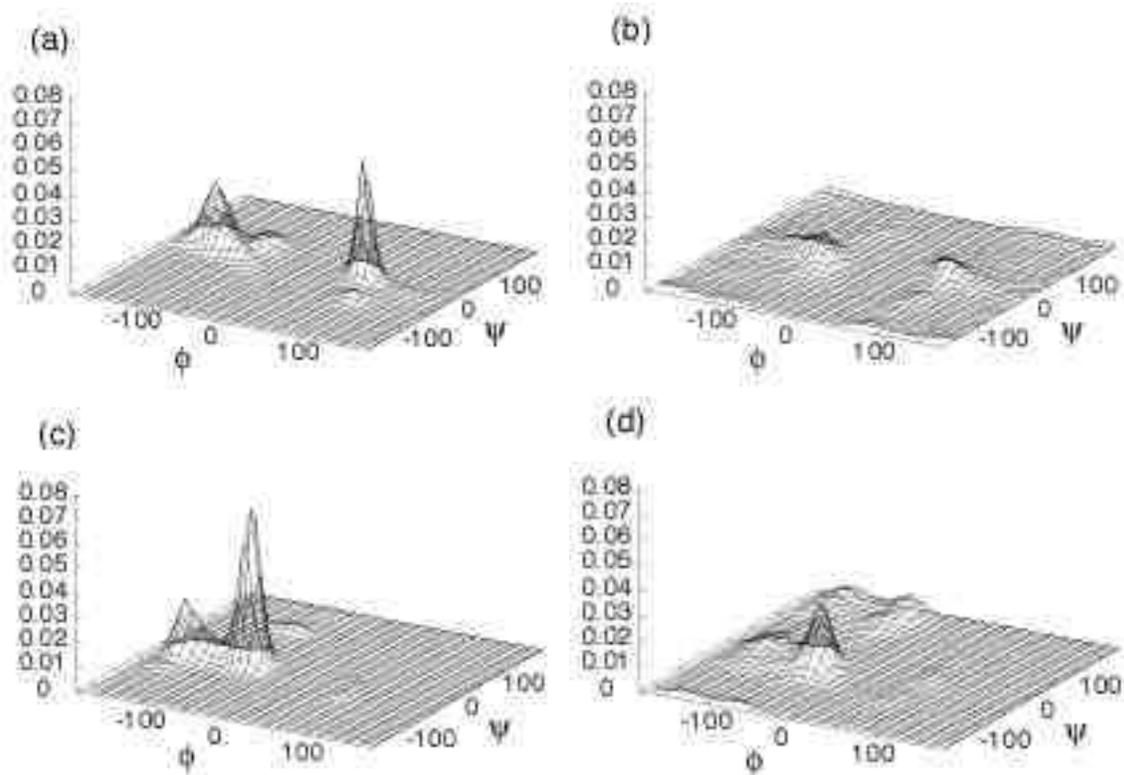

\begin{center}
\includegraphics[width=7.5cm,keepaspectratio]{oka_fig13a.epsf}
\includegraphics[width=7.5cm,keepaspectratio]{oka_fig13b.epsf}
\includegraphics[width=7.5cm,keepaspectratio]{oka_fig13c.epsf}
\includegraphics[width=7.5cm,keepaspectratio]{oka_fig13d.epsf}
\end{center}
\caption{Distributions of a pair of main-chain dihedral
angles $(\phi,\psi)$ of Met-enkephalin around
$T=300$ K for:
(a) Gly-2 in gas phase,
(b) Gly-2 in aqueous solution,
(c) Phe-4 in gas phase, and
(d) Phe-4 in aqueous solution.}
\label{fig13}
\end{figure}

\section{CONCLUSIONS}

In this article we reviewed uses of 
generalized-ensemble
algorithms for free-energy calculations in protein
folding.

We introduced two new generalized-ensemble
algorithms which are generalizations of the
replica-exchange method (REM) (we remark that
REM is also referred to as parallel tempering). 
The first one is the multidimensional
replica-exchange method (MREM), with which we
showed that the replica-exchange method
is not limited
to tempering (or temperature
exchange) and that we can also 
exchange parameters in the potential
energy.  One particular realization of this
method is replica-exchange umbrella sampling
(REUS) where we perform tempering and/or the
exchange of parameters that characterize
the umbrella potential.  
The second method is
the replica-exchange multicanonical
algorithm (REMUCA), in which we combine the
merits of REM and multicanonical algorithm (MUCA).  

With these new methods available,
we believe that we now have working simulation 
algorithms
which we can use for free-energy calculations
in protein folding.

\vspace{0.5cm}
\noindent
{\bf Acknowledgements}: \\
We are grateful to Dr. A. Kitao of Kyoto University
for useful discussions and collaboration.
Our simulations were performed on the Hitachi and other
computers at the Research Center for Computational
Science, Okazaki National Research Institutes.
This work is supported, in part, by a grant from 
the Research for the Future Program of the Japan Society for the 
Promotion of Science (JSPS-RFTF98P01101).\\

%%%%%%%%%%%%%%%%%%%%%%%%% references %%%%%%%%%%%%%%%%%%%

\end{document}